\documentclass[usenatbib]{mnras} 
\usepackage{graphicx}
\usepackage{amsmath}
\usepackage{array}
\usepackage{multicol}
\usepackage{multirow}
\usepackage{lscape}
\usepackage{rotating}
\usepackage{url}
\usepackage{color}
\usepackage{hyperref} 

\title[Spectroscopy of a gravitationally lensed SN Ia]{A spectroscopic look at the gravitationally lensed type Ia SN 2016geu at $z=0.409$}

\author[Cano et al.]{\noindent Zach Cano$^{1}$\thanks{zewcano@gmail.com}, Jonatan Selsing$^{2}$, Jens Hjorth$^{2}$, Antonio~de~Ugarte~Postigo$^{1,2}$, 
\newauthor
Lise Christensen$^{2}$, Christa Gall$^{2}$, D. A. Kann$^{1}$\\ 
\noindent $^{1}$Instituto de Astrof\'isica de Andaluc\'ia (IAA-CSIC), Glorieta de la Astronom\'ia s/n, E-18008, Granada, Spain. \\
\noindent $^{2}$Dark Cosmology Centre, Niels Bohr Institute, University of Copenhagen, Juliane Maries Vej 30, 2100 K\o benhaven \O{}, Denmark. \\
}

\begin{document}

\date{Accepted xx. Received xx; in original form xx}

\pagerange{\pageref{firstpage}--\pageref{lastpage}} \pubyear{2013}

\maketitle

\label{firstpage}

\begin{abstract}
The spectacular success of type Ia supernovae (SNe Ia) in SN-cosmology is based on the assumption that their photometric and spectroscopic properties are invariant with redshift.  However, this fundamental assumption needs to be tested with observations of high-$z$ SNe Ia.  To date, the majority of SNe Ia observed at moderate to large redshifts ($0.4 \le z \le 1.0$) are faint, and the resultant analyses are based on observations with modest signal-to-noise ratios that impart a degree of ambiguity in their determined properties.  In rare cases however, the Universe offers a helping hand: to date a few SNe Ia have been observed that have had their luminosities magnified by intervening galaxies and galaxy clusters acting as gravitational lenses.  In this paper we present long-slit spectroscopy of the lensed SNe Ia 2016geu, which occurred at a redshift of $z=0.409$, and was magnified by a factor of $\approx$55 by a galaxy located at $z=0.216$.  We compared our spectra, which were obtained a couple weeks to a couple months past peak light, with the spectroscopic properties of well-observed, nearby SNe Ia, finding that SN~2016geu's properties are commensurate with those of SNe Ia in the local universe.  Based primarily on the velocity and strength of the Si~\textsc{ii}~$\lambda$6355 absorption feature, we find that SN~2016geu can be classified as a high-velocity, high-velocity gradient and ``core-normal'' SN Ia.  The strength of various features (measured though their pseudo-equivalent widths) argue against SN~2016geu being a faint, broad-lined, cool or shallow-silicon SN Ia.  We conclude that the spectroscopic properties of SN~2016geu imply that it is a normal SN Ia, and when taking previous results by other authors into consideration, there is very little, if any, evolution in the observational properties of SNe Ia up to $z\approx 0.4$.
\end{abstract}

\begin{keywords}
supernovae: general; supernovae: individual: SN 2016geu; gravitational lensing: strong; cosmology: miscellaneous.
\end{keywords}

\begin{figure*}
 \centering  
 \includegraphics[bb=18 180 594 612,keepaspectratio=true, width=\hsize, trim={0 0 0 0}]{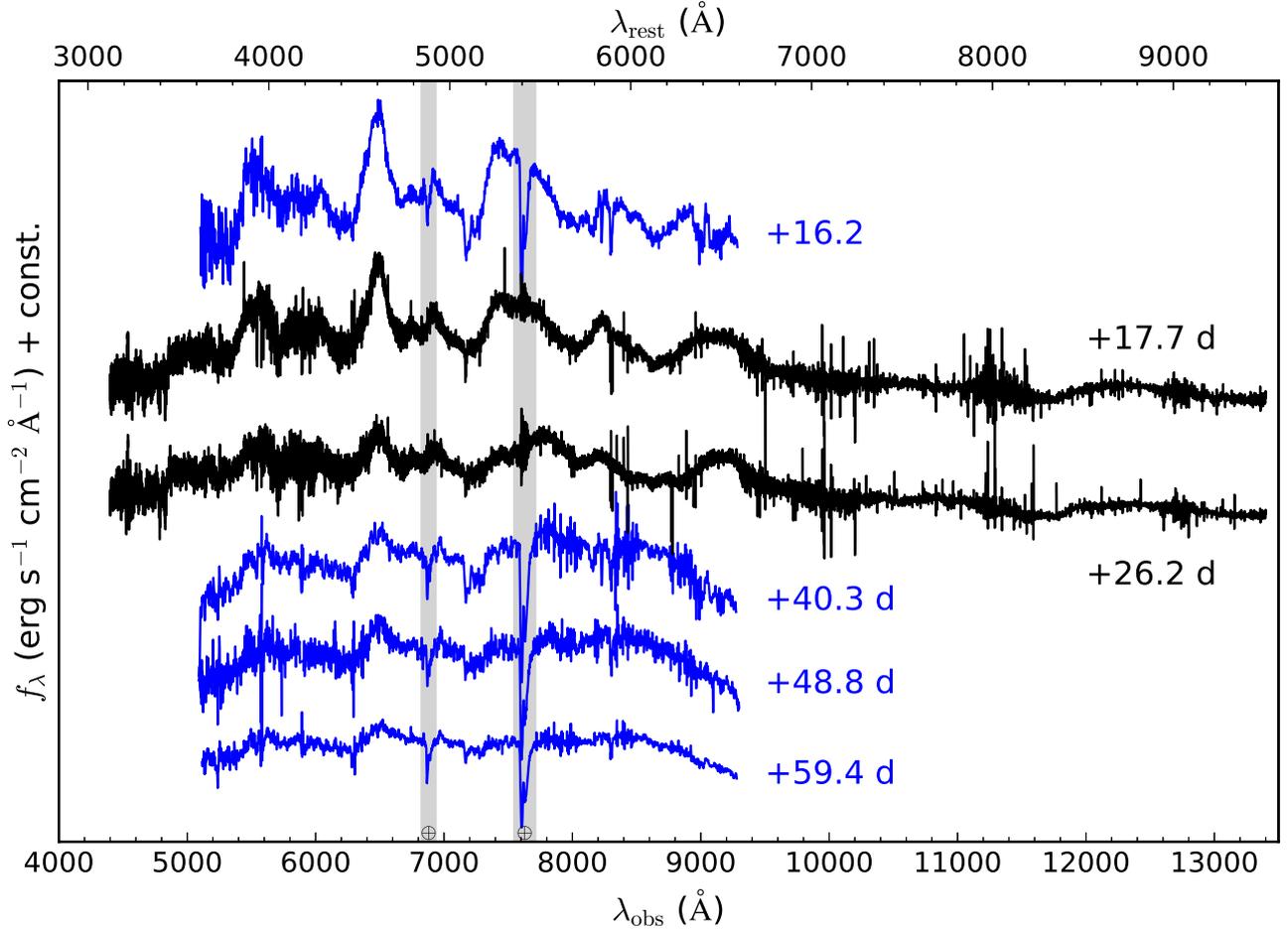}  
 \caption{Spectroscopic time-series of SN~2016geu obtained with GTC/OSIRIS (blue) and VLT/X-shooter (black), which have been dereddened for foreground and host extinction.  Contributions from the lensing and host galaxies have not been removed.  Observer-frame and rest-frame wavelengths are shown on the bottom and top $x$-axes, respectively.  Times (rest-frame) are given relative to peak $B$-band light ($t_{B}^{\rm max} = 2457654.6$).  Regions of contamination from telluric lines are denoted as the shaded-grey regions.  All spectra are in their native resolution, and some strong lines have been removed manually for presentation purposes.}
 \label{fig:1_spectra}
\end{figure*}

\section{Introduction}

The study of type Ia supernovae (SNe Ia) encompasses many fields of astrophysical research: they are connected with the end-points of stellar evolution in binary systems, and elements nucleosynthesised during their explosions, along with their enormous energy input, enrich the environments of their host galaxies.  They also play a central role in SN cosmology, and provide an extremely useful tool for determining cosmological distances. These in turn provide constraints on fundamental cosmological properties, including the expansion rate of the universe and the amount of matter and energy contained therein \citep{Riess98,Perlmutter99}.
 
The role of SNe Ia in cosmology cannot be understated.  The absolute luminosity and the width of their light-curves are correlated \citep{Phillips93}, which provides a means to reduce the amount of scatter in SNe Ia Hubble diagrams and provide better constraints on cosmological models.  Hence, two-parameter luminosity calibrations are currently used in SNe Ia research, which includes a colour term in the fitting process (e.g. \citealt{Tripp98,Guy07}), while additional techniques aimed at reducing the Hubble-diagram scatter are constantly being sought after (e.g. \citealt{Hoeflich10}).

Despite the success of SNe Ia as cosmological probes, fundamental understanding of why they are such good standardizable candles is generally lacking.  Sub-types of SNe Ia were identified as early as the mid-1980s (e.g. SN~1984A \citealt{Branch87,Phillips87}), and later in the early 1990s (e.g. SNe 1991bg and 1991T \citealt{Filippenko92a,Filippenko92b,Phillips92}).  Indeed the overall spectroscopic diversity of SNe Ia \citep{Filippenko97}, followed by the classification of a fraction of SNe Ia into sub-types (see Section \ref{sec:classifications}), prompts the question of just why do the bulk of observed SNe Ia, i.e. 'normal SNe Ia', have such ideal properties that facilitate their successful use in SN cosmology?  

Further uncertainties fuel this question: what are the progenitor systems of SNe Ia?  How do their environments affect their observed properties \citep{Hamuy95,Howell01,Kelly15a}?  Do all SNe Ia, and their corresponding sub-classes, appear the same across all of cosmic time, or is there a redshift evolution?  Certainly the physical and chemical properties (mass, age, metal and dust content) of galaxies change with time, and if the physical environments hosting SNe Ia directly affect their physical and observational properties, perhaps by influencing the amount of nickel synthesized during the explosion (e.g. \citealt{Hoeflich98,Timmes03}), this will have a direct influence on a given SN's luminosity.  Indeed, as current SN Ia cosmology is based on the assumption of \textit{no} evolution of the photometric properties with redshift, unambiguously addressing the latter question is of great importance to the field.

In this paper, we have attempted to address the question of redshift invariance of SNe Ia. Previous studies have tested this assumption using different approaches.  The seminal works of \citet{Riess98} and \citet{Perlmutter99}, who investigated SNe Ia up to $z=0.5$, found no evolution in their spectra and light-curves.  However, \citet{Riess99} compared a low-$z$ and high-$z$ sample of SNe Ia, and found the rise times of the two groups differed by more than $5\sigma$.  \citet{Hook05} studied a sample of $N=14$ SNe Ia in the redshift range $0.17\le z \le0.83$, and found that both the spectral time sequences and the measured Ca \textsc{ii} H\&K velocities were indistinguishable from low-$z$ SNe Ia.  Conversely, using the assumption that SNe Ia arise from two populations (prompt and delayed), \citet{Howell01} found that SNe Ia at redshifts up to $z=1.5$ may be, on average, 12\% more luminous than their low-$z$ counterparts.  Moreover, based on Hubble Space Telescope (\emph{HST}) spectroscopic observations of low-$z$ and high-$z$ SNe Ia ($0.4 \le z \le 0.9$), \citet{Maguire12} found a modest evolution in their NUV spectra ($\approx$ $3\sigma$ significance), where there appeared to be a UV-flux deficit in the more distant SNe Ia relative to the nearby sample.  Collectively, these studies may suggest that there is a small, yet appreciable, difference in the photometric and spectroscopic properties of nearby and high-$z$ SNe Ia, though this is far from being certain.


Many of the aforementioned studies relied on spectroscopic observations of high-$z$ SNe Ia, which are inevitably faint and suffer from low signal-to-noise ratios (SNRs) that limit the amount of unambiguous information that can be ascertained of these events.  Fortunately, Nature sometimes reveals glimpses of her secrets in beautiful and awe-inspiring ways.  Massive galaxies and galaxy clusters have been observed to act as weak and strong gravitational lenses, where they magnify distant objects that intersect the line-of-sight with themselves and Earth.  To date a few gravitationally lensed SNe have been observed, including the famous type II SN Refsdal \citep{Kelly15b}, which as predicted by \citet{Refsdal64} had multiple images of the same SN, where the time delay between them is proportional to the value of the Hubble constant.  Lensed SNe Ia have also been detected: PS1-10afx, at a redshift of $z=1.39$, was shown by \citet{Quimby13} to be more than 30 times brighter than expected for an unlensed SNe Ia at a comparable redshift.  \citet{Petrushevska17} recently demonstrated that the spectroscopic properties of PS1-10afx, in comparison to the median spectra of nearby SNe Ia, showed no evidence for spectral evolution between it and low-$z$ SNe Ia.  SN HFF14Tom ($z=1.3457$) was discovered behind the galaxy cluster Abell 2744 ($0.308$), where the authors demonstrated its use as a direct probe of galaxy lens models \citep{Rodney15}.

Here we report on the spectroscopic properties of another lensed SNe Ia, SN~2016geu, which occurred at a redshift of $z=0.409$ and was $\approx 55$ times brighter than an unlensed SNe Ia at the same distance.  Such an increase in the brightness of a SN Ia has provided us with an opportunity to investigate the spectroscopic properties of these objects at a redshift where such properties are poorly understood using the largest telescopes and the most advanced spectrographs currently available.  Our goal therefore was to determine its spectral properties as a function of time (where its photometric properties were determined elsewhere by \citealt{Goobar17}).  This paper is organised as: in Section \ref{sec:background} we present relevant background information of SN~2016geu published by \citet{Goobar17}.  In Section \ref{sec:data_reduction} we describe our data reduction procedures, while in Section~\ref{sec:lens_subtraction} we describe our method to account for and remove contaminating flux from the lensing galaxy in our spectra.  In Section~\ref{sec:SNIa_compare} we compare our spectroscopic time-series with nearby SNe Ia 1994D and 2011fe.  In Section \ref{sec:vel_pWs} we determine the pseudo-equivalent widths and line velocities of several transitions in our spectra, and compare them with those presented in the literature.  We sub-classify SN~2016geu in Section \ref{sec:classifications}, and compare our observations with synthetic SYN++ spectra in Section \ref{sec:synthetic}.  Finally in Section \ref{sec:conclusion} we discuss and summarise our conclusions.  All times and phases are given in the rest-frame unless specified otherwise, while $\left < \right >$ indicates an average value.

\begin{table}
\centering
\setlength{\tabcolsep}{4.0pt}
\setlength{\extrarowheight}{3pt}
\caption{SN 2016geu: vital statistics}
\label{table:vitals}
\begin{tabular}{rl}
\hline			
SN 2016geu	&	Ref.	\\
\hline			
RA(J2000) = 21$^{\rm h}$04$^{\rm m}$15.86$^{\rm s}$	&	G17	\\
Dec(J2000) = $-06^{\rm d}$20$^{\prime}$24.5$^{\prime \prime}$	&	G17	\\
$z_{\rm SN}=0.409$	&	G17	\\
$z_{\rm lens}=0.216$	&	G17	\\
$\mu_{\rm lens} =4.37\pm0.15$ &	G17	\\
$E(B-V)_{\rm total} = 0.31 \pm 0.05$ mag	&	G17	\\
$t_0 (JD) = 2457629.9 \pm 0.3$ & G17 \& this work	\\
$m_{B}^{\rm peak} = 18.42 \pm 0.22$	&	G17	\\
$t_{B}^{\rm max} (JD) = 2457654.6 \pm 0.2$	&	G17	\\
$\Delta m_{15, B} = 1.08\pm0.19$ 	&	G17 \& this work	\\
$v_{\rm Si~\textsc{ii}~\lambda6355}^{\rm max} = 11,950\pm140$~km~s$^{-1}$ & this work \\
$v_{10} = 10,850\pm250$~km~s$^{-1}$ & this work \\
$\dot{v} = -110.3\pm10.0$~km~s$^{-1}$~d$^{-1}$ & this work \\
\hline					
\end{tabular}
\begin{flushleft}
G17: \cite{Goobar17} \\
\end{flushleft}
\end{table}

\begin{figure*}
 \centering
 \includegraphics[bb=0 0 1440 720,scale=0.35,keepaspectratio=true]{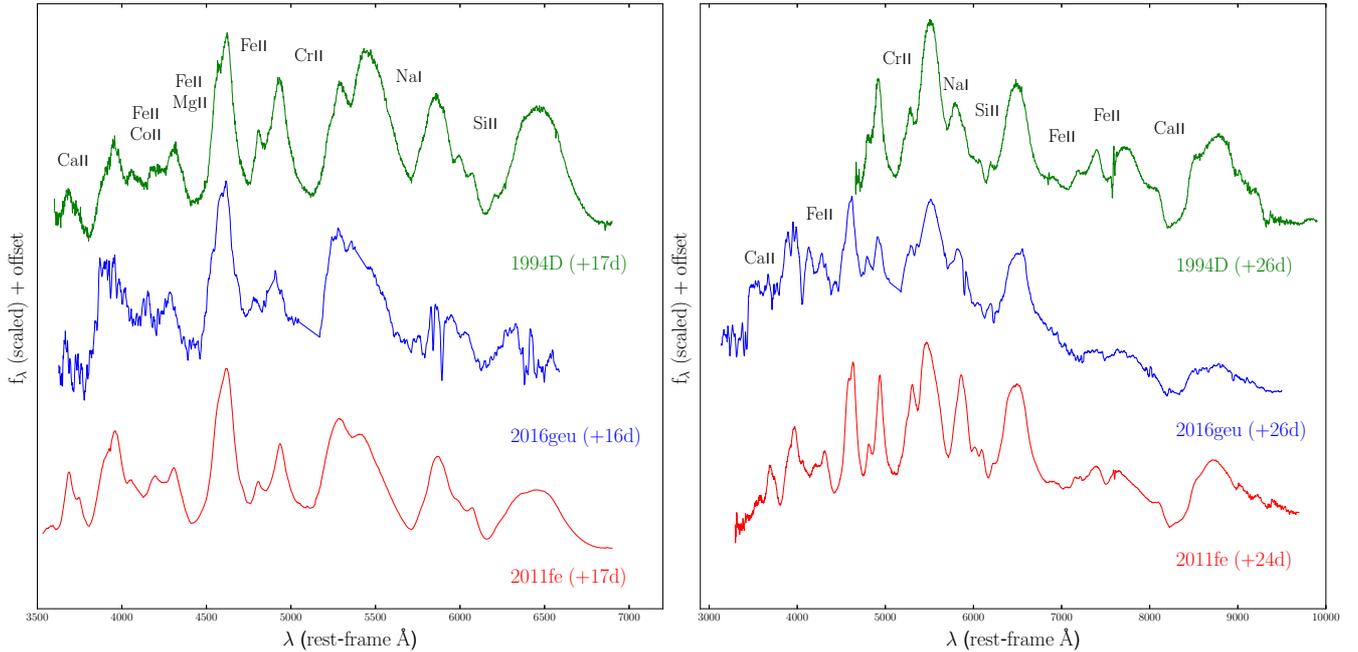}
 \caption{Comparison of the rest-frame spectra of SN 2016geu (which include contributions from the underlying host and lensing galaxy) with two nearby SNe Ia: SN 1994D (green) and SN 2011fe (red).  \textit{Left}: GTC spectrum obtained 16.2~days after peak $B$-band light. \textit{Right}: X-shooter spectrum obtained 26.2~days after peak $B$-band light.  Note that the X-shooter spectrum spans a wider wavelength range than the GTC one. The main absorption features seen in the comparison spectra are presented, which consist of iron-peak elements (Fe~\textsc{ii}, Cr~\textsc{ii} and Co~\textsc{ii}) and intermediate-mass elements (Ca~\textsc{ii}, Mg~\textsc{ii}, Na~\textsc{i}D and Si~\textsc{ii}).  Many of the features in the SN~2016geu spectra can also be attributed to these elements and ions. Both SN 2016geu spectra have been smoothed, and regions of telluric absorption (oxygen bands A \& B, and water) removed, for presentation purposes.}
 \label{fig:2_SN_compare}
\end{figure*}

\section{SN 2016geu - Background Information}
\label{sec:background}

SN~2016geu \citep{Goobar17} was discovered by the intermediate Palomar Transient Factory (iPTF) on 05 September, 2016 at RA = 21$^{\rm h}$04$^{\rm m}$15.86$^{\rm s}$ and Dec = $-06^{\rm d}$20$^{\prime}$24.5$^{\prime \prime}$ (J2000) in SDSS galaxy J210415.89-062024.7.  Spectroscopic classification of the optical transient was performed with the SED Machine \citep{Ritter14} on the Palomar 60~inch telescope (P60) on 02 October, 2016, where it was found to be a spectroscopically normal SN Ia at $z\approx0.4$.  At this redshift, its apparent $R$-band magnitude ($R=19.5$) was more than 3.5 magnitudes brighter than a normal SN Ia at the same redshift, which prompted the realization that its increased brightness may be due to an intervening galaxy acting as a gravitational lens.  From narrow Ca \textsc{ii} H\&K and Na~\textsc{i}D absorption features, as well as weak H$\alpha$ and [N\textsc{ii}] emission lines in the Palomar 200~inch telescope (P200) and the 2.2~m Nordic Optical Telescope (NOT) spectra, \citet{Goobar17} found the redshift of the SN and the lensing system to be $z=0.409$ and $z=0.216$, respectively.  In a Keck $K$-band adaptive-optics image obtained  on 13 October, 2016, multiple images of the SN were recorded, which consisted of a partial Einstein ring of the host galaxy and two point sources.  Additional \emph{HST} imaging reinforced this observation \citep{Goobar17}.  

Using SALT2 \citep{Guy07}, \citet{Goobar17} showed that SN~2016geu had a (corrected) peak $B$-band apparent magnitude of $m_{B}^{\rm corr} = 18.42 \pm 0.22$, which it reached on 2016 September 23.10 UT (in Julian date: $t_{B}^{\rm max} = 2457654.6 \pm 0.2$).  They also found the value of the time-stretch ($x_1$) parameter to be $0.08\pm0.19$, which using the relations from \citet{Guy07}, implies $\Delta m_{15,B} \approx 1.08$~mag, where $\Delta m_{15}$ is a measure of how much the light-curve dims from peak light to 15 days later.  Next, it was concluded that SN~2016geu was magnified by the lensing galaxy by $4.37\pm0.15$~magnitudes, and the SN was moderately reddened (the total line-of-sight extinction was found to be $E(B-V)=0.31\pm0.05$, for $R_{V}=3.1$, which is the value used in this work).  An in-depth discussion of the different sources of reddening, as determined from Na~\textsc{i}D modelling, is presented by \citet{Gall17}.  

For a set of 391 K-corrected $B$- and $V$-band SN Ia light-curves discovered by the Sloan Digital Sky Survey-II Supernova Survey, \citet{Hayden10} found an average rise-time to peak $B$-band light of $17.38\pm0.17$~days.  If we relate this rise-time to the time of peak $B$-band light found for SN~2016geu, it implies an explosion date of $t_0 \approx 2457630.1$ (2016 August 29.60 UT).  Next, \citet{Goobar17} noted the very similar light-curve evolution of SN~2016geu relative to SN~2011fe, the latter of which had a rise-time of $17.69$~days.  If we assume an identical rise time between the two SNe, it implies an explosion date for SN~2016geu of $t_0 \approx 2457629.7$ (2016 August 29.20 UT), which is about 10~hrs earlier than the previous estimate.  Combined, the average of these two is $t_0 = 2457629.9 \pm 0.3$, which is the value used throughout this paper.


\begin{table*}
\centering
\setlength{\tabcolsep}{6.0pt}
\setlength{\extrarowheight}{3pt}
\caption{SN 2016geu: Spectroscopy log}
\label{table:spectra_obs_log}
\begin{tabular}{ccccccc}
\hline													
UT Date		&	$t-t_0$ (d)$^\dagger$	&	$\Delta t_{B}^{\rm max}$ (d)$^\ddagger$	&	wavelength range (\AA)	&	Telescope \& Instrument	&	Grism		&	Exposure (s)	\\
\hline													
15 Oct 2016	&	33.7	&	+16.2	&	$5100-10,000$	&	GTC-OSIRIS	&	R1000R	&	$5 \times 300$	\\
18 Oct 2016	&	35.2	&	+17.7	&	$3200-22,000$	&	VLT-X-shooter	&	UVB+VIS+NIR	&	$4 \times 1184$	\\
30 Oct 2016	&	43.8	&	+26.2	&	$3200-22,000$	&	VLT-X-shooter	&	UVB+VIS+NIR	&	$4 \times 1184$	\\
18 Nov 2016	&	57.8	&	+40.3	&	$5100-10,000$	&	GTC-OSIRIS	&	R1000R	&	$3 \times 900$	\\
30 Nov 2016	&	66.3	&	+48.8	&	$5100-10,000$	&	GTC-OSIRIS	&	R1000R	&	$3 \times 1200$	\\
15 Dec 2016	&	77.0	&	+59.4	&	$5100-10,000$	&	GTC-OSIRIS	&	R1000R	&	$4 \times 1200$	\\
\hline													
\end{tabular}
\begin{flushleft}
All times and phases are given in the rest-frame of SN~2016geu ($z=0.409$).\\
$^\dagger$Phase relative to the estimated time of explosion ($t_0 = 2457629.9 \pm 0.3$).\\
$^\ddagger$Phase relative to peak $B$-band light ($t_{B}^{\rm max}=2457654.6\pm0.2$, see \citealt{Goobar17}). \\
\end{flushleft}
\end{table*}

\section{Data reduction}
\label{sec:data_reduction}

We obtained four epochs of spectroscopy of SN~2016geu with GTC-OSIRIS as part of the GTC DDT proposal GTC2016-058 (PI: Z. Cano) using grism R1000R.  The GTC spectra were reduced using standard techniques with \texttt{IRAF}-based scripts.  The reduced GTC spectra were flux calibrated with the $r$-band acquisition images obtained each night, where the latter were calibrated to two dozen SDSS standard stars in the SN field-of-view via a zeropoint, and the AB flux density was converted from units of erg~s$^{-1}$~cm$^{-2}$~Hz$^{-1}$ into units of erg~s$^{-1}$~cm$^{-2}$~\AA$^{-1}$ using an effective wavelength for the GTC/OSIRIS $r$-band filter of $\lambda_{\rm eff} = 6122.3$~\AA.  A normalization factor between the flux density of the $r$-band acquisition image and the GTC/OSIRIS spectrum was manually determined and applied to the latter.  Two spectra were obtained with the X-shooter instrument \citep{Vernet11} mounted on Unit Telescope 2 (UT2, Kueyen) of the Very Large Telescope (VLT) at the Paranal Observatory as part of ESO Program ID 098.A-0648(A) (PI: J. Hjorth), which were reduced using the VLT/X-shooter pipeline version 2.8.4 \citep{Modigliani10} run in physical mode along with dedicated \texttt{PYTHON} scripts for the post-processing of the images (see \citealt{Gall17} for a full description of the X-shooter data reduction and calibration procedure).  All the absolute-flux-calibrated spectra were then corrected for line-of-sight extinction, i.e. foreground Milky-Way extinction \citep{Schlegel98, SF11}, and rest-frame extinction \citep{Goobar17}.  A summary of all spectra used in this current analysis is given in Table~\ref{table:spectra_obs_log}, while the complete spectroscopic time-series is presented in Fig.~\ref{fig:1_spectra}.





\section{Lens galaxy subtraction}
\label{sec:lens_subtraction}

As the spectroscopic slit covered parts of SN~2016geu, as well as its host galaxy and the lensing galaxy, (both of which are thought to be elliptical galaxies; \citealt{Goobar17}), the obtained spectra consist of integrated light from three distinct components at two different redshifts. Therefore, in order to correctly investigate the pseudo equivalent widths (\textit{p}W) of SN~2016geu (Section \ref{sec:vel_pWs}), the spectra need to be cleaned of these contaminating sources. To accomplish this, the ``colour matching'' technique outlined in \citet{Foley12} was employed. As the relative flux contributions and spectral shapes of the lensing galaxy, host galaxy, and the SN are unknown in the spectra, the optimal approach is to make the lens- and host-subtracted spectra match the SN colours, the latter of which are obtained through difference imaging. For consistency, the spectra have all been scaled to the acquisition camera magnitudes inferred using an aperture covering the entire source.  Additionally, we used a Mannucci elliptical template \citep{Mannucci01,Mannucci05} to represent the host and lensing galaxies.

To do the colour matching, we determined the SN colours at each epoch by fitting the light curves published by \citet{Goobar17} with the SALT2 light curve fitting tool \citep{Guy07}, which were executed using SNCosmo \citep{Barbary16} and the best-fitting light curves were obtained through the use of \texttt{emcee} \citep{FM13}.  The best-fitting model parameters are: $t_0 = 57654.20 _{-0.20} ^{+0.20}$, $x_0 \cdot 10^{4}= 3.96 _{-0.10} ^{+0.10}$, $x_1 = 0.15 _{-0.20} ^{+0.21}$, and $\mathcal{C} = 0.25 _{-0.04} ^{+0.04}$. We corrected for Milky Way dust extinction using the dust maps of \citet{Schlegel98,SF11}.  These derived values are entirely consistent with those obtained by \citet{Goobar17} (see, e.g. Section \ref{sec:background}).

Equipped with the SALT2 model light curve fits, we calculated equivalent synthetic magnitudes from each spectroscopic observation using the SDSS $ugriz$ filter transmission curves\footnote{\url{http://www.sdss.org/instruments/camera/\#Filters}} and the two X-shooter spectra (as they span the widest wavelength range and hence more photometric bands that can be used to constrain the relative contribution of the lens and host). For both X-shooter observations, the host contamination contribution was found to be negligible, compared to the lens. However, as we see the Ca H \& K absorption lines from the host galaxy, and host flux is clearly visible in the adaptive optics, $K$-band Keck/Osiris observations presented in \citet{Goobar17} (their fig. 3), we know that there is some host contamination, but the colour matching is best achieved with a single template (i.e. lens galaxy only).   Hence, the derived \textit{p}Ws (Section \ref{sec:vel_pWs}) and comparison with synthetic spectra (Section \ref{sec:synthetic}), have had the lensing galaxy contribution removed, however there is an unknown contribution from the underlying host galaxy.  As such, strictly speaking the \textit{p}Ws should be considered as upper limits, as any additional host-template spectra removal will change the continuum levels of the spectra.

\section{Comparison with other SNe Ia}
\label{sec:SNIa_compare}

To date, several nearby SNe Ia have been detected within half a dozen Mpc.  Their close proximity has led to the acquisition of exquisite high-cadence and excellent SNR photometric and spectroscopic datasets.  One such event, which is now widely considered to be the archetype SN Ia, is SN~2011fe, which exploded in the Pinwheel Galaxy ($D=6.4$~Mpc, \citealt{Paturel03}), and it was shown to suffer from very little, if any reddening in its host galaxy \citep{Pereira13}.  Even closer SNe Ia have been detected, including SN~2014J (which exploded in M82, at a distance of $D=3.5$~Mpc), however unlike SN~2011fe, SN~2014J suffered from a large amount of extinction ($E(B-V)=1.22\pm0.5$~mag), and like other highly reddened SNe Ia (e.g. SN~2006X, \citealt{Wang08}), had a reddening law of $R_{V} = 1.4\pm0.1$ \citep{Goobar14,Amanullah14}.


In this work we compared our GTC and X-shooter spectroscopic observations of SN~2016geu with those of two nearby, well observed SNe Ia: SN~1994D \citep{Patat96}, which exploded in NGC~4526 (distance of $16.9\pm1.6$~Mpc), and SN~2011fe \citep{Pereira13}.  These datasets provide us with a tool to determine the spectral phase and a first-order approximation of the chemical content of the ejecta of SN~2016geu via direct comparison.  The spectra of the comparison SNe Ia were dereddened for foreground \citep{SF11} (assuming $R_{V} = 3.1$) and rest-frame extinction using the values from the aforementioned papers. The rest-frame reddening was applied to rest-frame spectra ($z=0.001494$ and $0.00080$, respectively).  We compared the spectra of SN~2016geu from two (rest-frame) epochs: the first GTC spectrum obtained at $t_{B}^{\rm max} = +16.2$~days, and the second X-shooter spectrum obtained at $t_{B}^{\rm max} = +26.2$~days, which are shown in Fig. \ref{fig:2_SN_compare}.

Visual inspection of the first GTC spectrum reveals several typical transitions seen in SN Ia at a couple weeks past peak light (e.g. \citealt{Parrent14}), including blended iron-peak elements (Fe~\textsc{ii}, Cr~\textsc{ii} and Co~\textsc{ii}), and several intermediate mass elements (Ca~\textsc{ii}, Mg~\textsc{ii}, Na~\textsc{i}D and Si~\textsc{ii}).  The strong absorption feature near 6150~\AA, which is likely blueshifted Si~\textsc{ii} $\lambda$6355, is clearly present in the first GTC spectrum (+16.2~days) and the two X-shooter spectra (+17.7 and +26.2~days), but absent in the latter three (later than +40.3~days), implying that the layers probed by the photosphere at these epochs are devoid of this element.  


\begin{figure*}
 \centering
 \includegraphics[bb=0 0 1440 720,scale=0.35,keepaspectratio=true]{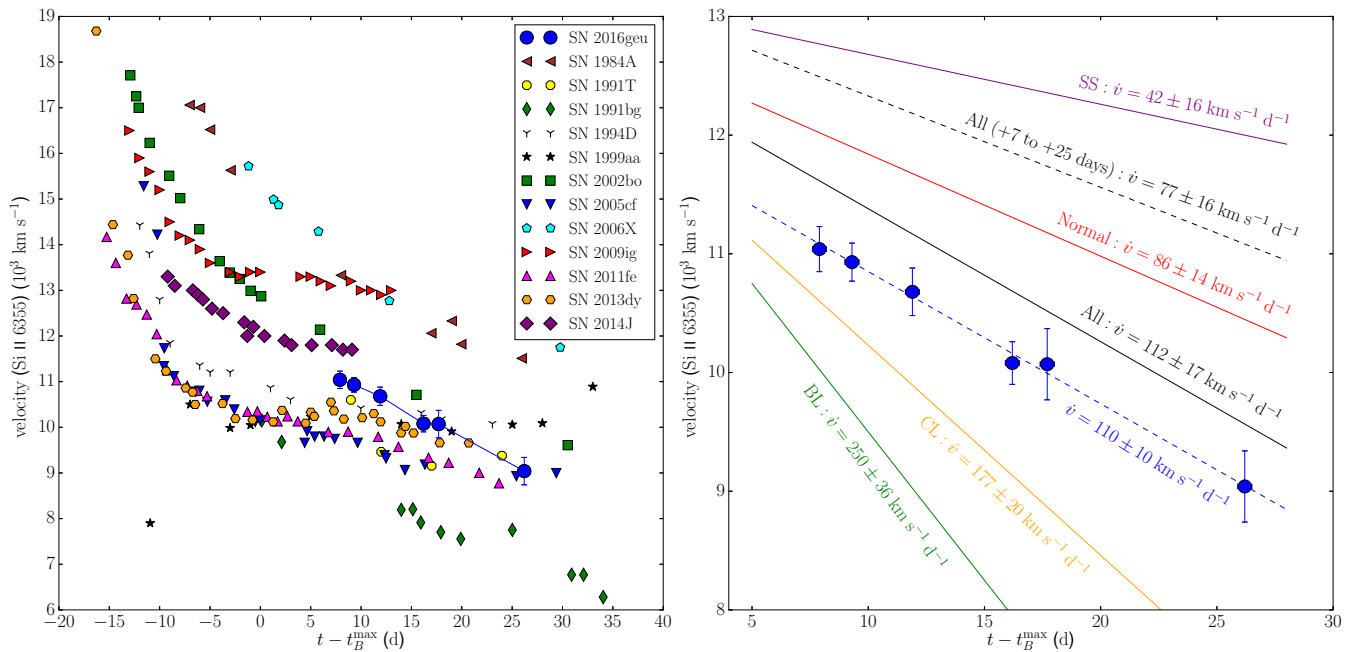}
 \caption{Blueshifted velocity of the Si~\textsc{ii}~$\lambda$6355 transition.  \textit{Left}: A comparison with other well-observed SNe Ia.  \textit{Right}: The slope of the velocity evolution with time for SN~2016geu is $\dot{v} = 110.3\pm10.0$~km~s$^{-1}$~d$^{-1}$, which was fit to the data from $+7.9 < t_{B}^{\rm max} < +26.2$ days (i.e. data presented here and those in \citealt{Goobar17}).  Plotted are the slopes, near peak $B$-band light, determined for the four SN Ia subclasses studied by \citet{Folatelli13}: Normal, broad-lined (BL), CL (cool), shallow-silicon (SS), and all classes combined.  Also plotted is the slope determined from data taken after peak $B$-band light ($+7 < t_{B}^{\rm max} < +25$ days), which matches well the phases investigated here for SN~2016geu.  It is seen that the slope in $\dot{v}$ for SN~2016geu is steeper than the combined (all), post-maximum sample of \citet{Folatelli13} by $\approx 2\sigma$. }
 \label{fig:3_v_Si}
\end{figure*}

\begin{figure}
 \centering
 \includegraphics[bb=0 0 576 432,scale=0.43,keepaspectratio=true]{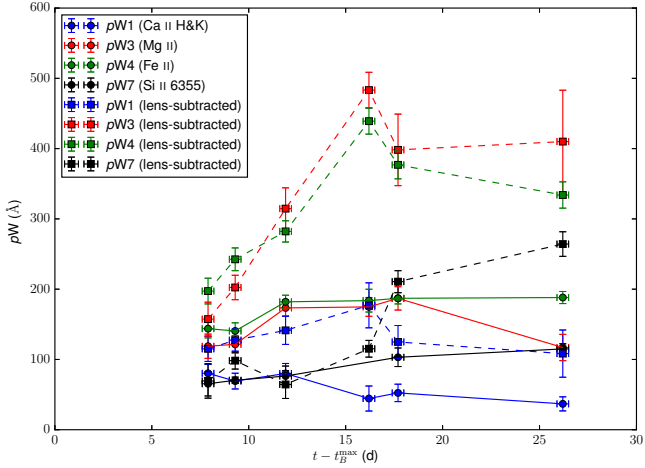}
 \caption{Evolution of the pseudo-equivalent width features with time: \textit{p}W1 (Ca~\textsc{ii} H\&K, in blue), \textit{p}W3 (Mg~\textsc{ii}, in red), \textit{p}W4 (Fe~\textsc{ii}, in green), and \textit{p}W7 (Si~\textsc{ii}~$\lambda$6355, in black).  Values derived from the spectra containing flux from the SN+host+lens are shown as filled circles and solid lines, while those obtained from the lens-subtracted spectra are shown as filled squares and dashed lines.}
 \label{fig:4_pWs}
\end{figure}

\begin{table*}
\centering
\setlength{\tabcolsep}{6pt}
\setlength{\extrarowheight}{3pt}
\caption{SN 2016geu: Line velocities and (pseudo) equivalent widths}
\label{table:linevels}
\begin{tabular}{ccccccccc}
\hline	
Phase$^{\dagger}$	&		$v_{\rm Si\lambda6355}$ ($10^3$~km~s$^{-1}$)				&		\textit{p}W1 (\AA)				&		\textit{p}W3 (\AA)				&		\textit{p}W4 (\AA)				&		\textit{p}W7 (\AA)				&	Telescope	&	Spectra Ref.	\\
\hline																																			
+7.9	&	$	11.04	\pm	0.19	$	&	$	80.3	\pm	12.8	$	&	$	118.8	\pm	17.2	$	&	$	143.8	\pm	13.1	$	&	$	65.4	\pm	17.4	$	&	P200	&	SN+host+lens	&	G17	\\
+7.9	&	$	-			$	&	$	115.1	\pm	17.9	$	&	$	157.4	\pm	24.1	$	&	$	197.3	\pm	18.3	$	&	$	69.4	\pm	24.5	$	&	P200	&	SN+host	&	G17	\\
+9.3	&	$	10.93	\pm	0.16	$	&	$	69.2	\pm	11.2	$	&	$	121.5	\pm	12.4	$	&	$	140.5	\pm	11.6	$	&	$	70.4	\pm	4.5	$	&	P200	&	SN+host+lens	&	G17	\\
+9.3	&	$	-			$	&	$	127.5	\pm	15.7	$	&	$	202.4	\pm	17.4	$	&	$	242.5	\pm	16.2	$	&	$	98.3	\pm	12.0	$	&	P200	&	SN+host	&	G17	\\
+11.9	&	$	10.68	\pm	0.20	$	&	$	80.0	\pm	14.3	$	&	$	173.3	\pm	11.5	$	&	$	182.0	\pm	9.4	$	&	$	76.4	\pm	14.2	$	&	NOT	&	SN+host+lens	&	G17	\\
+11.9	&	$	-			$	&	$	141.4	\pm	20.0	$	&	$	314.6	\pm	29.6	$	&	$	282.2	\pm	15.1	$	&	$	64.4	\pm	19.9	$	&	NOT	&	SN+host	&	G17	\\
+16.2	&	$	10.08	\pm	0.18	$	&	$	44.4	\pm	17.8	$	&	$	174.8	\pm	13.4	$	&	$	183.7	\pm	16.2	$	&	$	>39.6			$	&	GTC	&	SN+host+lens	&	this work	\\
+16.2	&	$	-			$	&	$	177.0	\pm	32.0	$	&	$	483.3	\pm	25.3	$	&	$	439.1	\pm	18.5	$	&	$	115.2	\pm	11.9	$	&	GTC	&	SN+host	&	this work	\\
+17.7	&	$	10.07	\pm	0.30	$	&	$	52.4	\pm	12.4	$	&	$	186.9	\pm	16.6	$	&	$	187.0	\pm	8.0	$	&	$	103.1	\pm	13.2	$	&	XS	&	SN+host+lens	&	this work	\\
+17.7	&	$	-			$	&	$	124.9	\pm	23.4	$	&	$	398.2	\pm	50.9	$	&	$	376.9	\pm	19.9	$	&	$	210.7	\pm	15.6	$	&	XS	&	SN+host	&	this work	\\
+26.2	&	$	9.04	\pm	0.30	$	&	$	36.8	\pm	10.0	$	&	$	117.0	\pm	18.7	$	&	$	188.1	\pm	8.6	$	&	$	115.0	\pm	7.3	$	&	XS	&	SN+host+lens	&	this work	\\
+26.2	&	$	-			$	&	$	108.3	\pm	33.6	$	&	$	410.0	\pm	73.1	$	&	$	334.0	\pm	18.7	$	&	$	264.2	\pm	17.5	$	&	XS	&	SN+host	&	this work	\\
\hline					
\end{tabular}
\begin{flushleft}
$^{\dagger}$ Phase relative to peak $B$-band light.
\end{flushleft}
\end{table*}

\section{Line velocities and (pseudo) equivalent widths}
\label{sec:vel_pWs}

\subsection{Line velocities}

Under the assumption that the absorption feature near 6150~\AA~is unblended Si~\textsc{ii}~$\lambda$6355, we fit it with a single Gaussian profile to determine its minimum, and hence its blue-shifted velocity.  Using a  method employed in previous works \citep{Cano14,Cano15,Cano17}, we used \texttt{python} to do the fitting, and employed a bootstrap (i.e. a Monte Carlo sampling and replacement) routine to estimate the errors in the local minima by utilizing the error spectrum from each epoch.  In this manner both the statistical and flux errors were appropriately considered.  Systematic errors will undoubtedly be introduced from the selection of the \textit{p}W end-points, and we have endeavored to be as consistent as possible with the spectra obtained with different instruments.  Nevertheless, such errors are likely present and may be under-represented in the quoted \textit{p}W errors.  The bootstrap-fitting procedure was performed on our GTC and X-shooter spectra, as well as those published in \citet{Goobar17}, which where retrieved from the Weizmann Interactive Supernova data REPository\footnote{\url{https://wiserep.weizmann.ac.il/}} (WISeREP, \citealt{YaronGY12}).  All spectra were corrected for foreground and host extinction and converted to rest-frame wavelengths.  A plot of the expansion velocity of Si~\textsc{ii}~$\lambda$6355 for SN~2016geu is presented in Fig.~\ref{fig:3_v_Si}, and the values can be found in Table \ref{table:linevels}.  Data of the comparison SNe Ia, which were chosen to represent the various sub-types of SNe (i.e. normal, faint, high-velocity, etc.; see Section \ref{sec:classifications}), are from \citet{Phillips92,Garavini04,Wang09,Marion13,Marion15,Pan15}, and references therein.

Next, we determined the rate of change of the Si~\textsc{ii}~$\lambda$6355 expansion velocity by fitting a straight line to the data to determine its slope ($\dot{v}$) and $y$-intercept (i.e. the expansion velocity at peak $B$-band light).  The errors were determined using an additional bootstrap (i.e. Monte Carlo sampling and replacement using the error spectra) analysis.  The best-fitting values are: $\dot{v} = -110.3 \pm 10.0$~km~s$^{-1}$~d$^{-1}$ and $v_{B}^{\rm max} = 11,950 \pm 140$~km~s$^{-1}$.  The velocity at $t_{B}^{\rm max} + 10$~days is $v_{10} = 10,850 \pm 250$~km~s$^{-1}$.  We note that the inferred velocity at peak $B$-band light is calculated from the assumption that the Si~\textsc{ii}~$\lambda$6355 expansion velocity evolves linearly for times from peak light and later (i.e. we have simply extrapolated to $t_{B}^{\rm max}$ using the linear fit). Inspection of the Si~\textsc{ii}~$\lambda$6355 velocity evolution of the comparison SNe Ia in Fig. \ref{fig:3_v_Si} reveals that while some do evolve linearly from peak $B$-band light onwards, e.g. SN 2002bo, which is a HV SN Ia, (see Section \ref{sec:classifications}), and SNe 1994D and 2005cf, which are normal SNe Ia, others do not (e.g. SNe 1984A and 2006X, both HV SNe Ia).

\subsection{Pseudo equivalent widths}

The expansion velocity, and its rate of change with time, can be used to sub-classify a given SN Ia (see Section \ref{sec:classifications}). Another useful classification diagnostic is the strength of various transitions in the optical spectra of SNe Ia, which are inferred by the equivalent width of one or more transitions.  In this work we have adopted the \textit{p}Ws defined by \citet{Folatelli04}, and used in several previous works (e.g. \citealt{Branch06,Garavini07,Folatelli13}).  In total, \citet{Folatelli04} defined eight different regions occurring between the near-UV and near-IR Ca~\textsc{ii} features.  All eight regions are usually present near peak $B$-band light, however by two weeks after peak light, three regions defined by different Si~\textsc{ii} transitions ($\lambda$4130, $\lambda\lambda$5449,5622, and $\lambda$5972) have disappeared.  In our spectra, as well as those from \citet{Goobar17}, four \textit{p}W regions are visible: \textit{p}W1 (Ca~\textsc{ii} H\&K), \textit{p}W3 (Mg~\textsc{ii}), \textit{p}W4 (Fe~\textsc{ii}), and \textit{p}W7 (Si~\textsc{ii}~$\lambda$6355).

We adopted the fitting procedure outlined in \citet{Garavini07}, which involves defining a pseudo continuum and finding the \textit{p}W between hand-chosen limits (we adhered to the blueward and redward limits defined in \citealt{Folatelli13}, their table 5 when choosing the limits of each \textit{p}W feature).  Our \textit{p}W values are presented in Fig. \ref{fig:4_pWs} and Table \ref{table:linevels}, which were derived for the spectra containing light from (1) the SN, its host and the lensing galaxy, and (2) just the SN and its host (see Section \ref{sec:lens_subtraction}).  In comparison with the sample of $N=93$ SNe Ia observed by the Carnegie Supernova Project (CSP) presented by \citet{Folatelli13}, the general evolution of the \textit{p}Ws of SN~2016geu follow those in the CSP sample (see figs. 10 \& 11 in \citealt{Folatelli13}).  In the CSP sample, near peak $B$-band light, the \textit{p}W1 values have a very large swath of values ($70-250$~\AA), which cluster more closely together near 100~\AA~after $2-4$~weeks.  There is a slight hint that the \textit{p}W1 values decrease in time. For SN~2016geu, the \textit{p}W1 values of the SN+host+lens slightly decrease over the time of the observations from 80~\AA~to 40~\AA, but with large errorbars.  For the lens-subtracted spectra, the initial and final values are around 110$\sim$120~\AA, and reach a maximum of almost 180~\AA~ in the GTC spectrum at $t_{B}^{\rm max} + 16$~days.

The \textit{p}W3 (Mg~\textsc{ii}) values in the CSP sample also have a wide range of values at all epochs (relative to $B$-max), but most have a value near \textit{p}W3 $\approx100$~\AA~ near peak $B$-band light, and a sharp increase to $\approx250-300$~\AA~between 5 and 15 days post-peak.  Similar behaviour is seen here for SN~2016geu, where the SN+host+lens \textit{p}W3 values increase from $\approx 120$~\AA~ to $\approx170$~\AA~ between 9.3~days and 11.9~days past $B$-max. For the lens-subtracted spectra, the \textit{p}W3 values increase steadily from 160$\sim$480~\AA~ from $t_{B}^{\rm max} + 7.9$ to $+16.2$~days, and level out around 400~\AA~ in the final two epochs.  We note that the \textit{p}W3 values from the GTC and X-shooter spectra taken two days apart differ by about 85~\AA, where the GTC spectrum gives larger values.  Interestingly, in the final X-shooter spectrum, the values for the lens-subtracted/unsubtracted spectra differ by a factor of almost four, meaning that the removal of the elliptical lens template was vital for properly charting the evolution of this particular pseudo-equivalent width.

Next, the \textit{p}W4 values also evolve in tandem with the \textit{p}W3 values, for both the lens-subtracted and -unsubtracted spectra, and they do so at a rate similar to that seen in the CSP sample.  Finally, the Si~\textsc{ii}~$\lambda$6355 feature (\textit{p}W7), is also seen to increase in the CSP sample in the 20 days following peak $B$-band light, increasing from $\approx 100$~\AA~ to $\approx 200$~\AA~ (with large scatter) during this time.  For SN~2016geu, the value of \textit{p}W7 also increases with time, though its value at all epochs appears to occur at the lower end of the \textit{p}W7 distribution seen in the CSP sample for the unsubtracted spectra, while the lens-subtracted values are generally larger and more closely match the behaviour seen in the CSP sample.

\section{Classifying SN 2016geu}
\label{sec:classifications}

In order to better understand the diversity of SNe Ia, and hence facilitate their use as standardizable candles with ever-improving precision, several statistical studies have been performed.  The results of these studies are sub-classifications of SNe Ia, based primarily on different spectral features, including blueshifted velocities and equivalent widths of several Ca, Si, Mg and Fe transitions.  

\subsection{Benetti et al. (2005)}

\citet{Benetti05} investigated a sample of $N=26$ SNe Ia, including SNe 1991T and 1991bg, and found three general sub-classes: FAINT, high velocity gradient (HVG) and low velocity gradient (LVG).  In their classification scheme, the primary indicator was the Si~\textsc{ii}~$\lambda$6355 transition velocity, as well as its rate of change  with time ($\dot{v}$).  FAINT SNe Ia had peak magnitudes commensurate with that of SN~1991bg ($\left < M_B \right > = -17.2$, $\sigma=0.6$~mag), low expansion velocities ($v_{10} = 9200$~km~s$^{-1}$, $\sigma = 600$~km~s$^{-1}$) and rapid velocity gradients ($\left < \dot{v} \right > = -87$~km~s$^{-1}$~d$^{-1}$, $\sigma=20$~km~s$^{-1}$~d$^{-1}$), and quickly evolving light-curves ($\left < \Delta m_{15, B} \right > = 1.83$~mag, $\sigma=0.09$~mag).  HVG SNe Ia, as their name suggests, have large expansion velocities ($v_{10} = 12,200$~km~s$^{-1}$, $\sigma = 1100$~km~s$^{-1}$) and gradients ($\left < \dot{v} \right > = -97$~km~s$^{-1}$~d$^{-1}$, $\sigma=16$~km~s$^{-1}$~d$^{-1}$), while  those of the LVG group have much slower velocity gradients ($\left < \dot{v} \right > = -37$~km~s$^{-1}$~d$^{-1}$, $\sigma=18$~km~s$^{-1}$~d$^{-1}$) and intermediate expansion velocities ($v_{10} = 10,300$~km~s$^{-1}$, $\sigma = 300$~km~s$^{-1}$).  The latter two groups have similar absolute, peak $B$-band magnitudes ($\left < M_B \right > = -19.3$, $-19.3$, respectively) and similar light-curve widths ($\left < \Delta m_{15, B} \right > = 1.2$, $1.1$~mag, respectively).

In the \citet{Benetti05} classification scheme, SN~2016geu has a light-curve width ($\left < \Delta m_{15, B} \right > = 1.1\pm0.2$~mag) similar to the averages of the HVG and LVG groups, but is more than 8$\sigma$ smaller than the average of the FAINT group.  In terms of the $v_{10}$ diagnostic, the value measured for SN~2016geu is almost 3$\sigma$ larger than the FAINT average, and within 1$\sigma$ and 2$\sigma$ of the HVG and LVG group averages, respectively.  Finally, in terms of the velocity gradient, SN~2016geu is consistent with both the FAINT and HVG group, but is more than 4$\sigma$ larger than the average of the LVG group.  Of all three groups, SN~2016geu is most consistent with the HVG group, and quite inconsistent with the FAINT group.

\subsection{Branch et al. (2006)}

Building upon the classification scheme devised by \citet{Folatelli04}, and extended by several authors \citep{Garavini07,Blondin12}, and notably \citet{Branch06}, \citet{Folatelli13} analysed a spectroscopic dataset of 604 spectra of $N=93$ SNe Ia that spanned a range of $-12$ to $+150$ days relative to peak $B$-band light.  Based on the strengths of the \textit{p}W6 and \textit{p}W7 (Si \textsc{ii} $\lambda$5972 and $\lambda$6355, respectively; see Section \ref{sec:vel_pWs}), four sub-classifications were devised: (i) cool (CL): \textit{p}W6 $>$ 30~\AA; (ii) broad-lined (BL): \textit{p}W6 $<$ 30~\AA~and \textit{p}W7 $>$ 105~\AA; (iii) shallow-silicon (SS): \textit{p}W7 $<$ 70~\AA; and (iv) core-normal (CN): \textit{p}W6 $\le$ 30~\AA~and 70 $\le$ \textit{p}W7 $\le$ 105~\AA.

Furthermore, the Si~\textsc{ii}~$\lambda$6355 velocity evolution of the four sub-classes differ (near peak $B$-band light), as depicted in Fig. \ref{fig:3_v_Si}: (i) CL: $-177\pm20$~km~s$^{-1}$~d$^{-1}$; (ii) BL: $-250\pm36$~km~s$^{-1}$~d$^{-1}$; (iii) SS: $-42\pm16$~km~s$^{-1}$~d$^{-1}$; and (iv) CN: $-86\pm14$~km~s$^{-1}$~d$^{-1}$.  The combined value for all sub-classes near peak $B$-band light was $-112\pm17$~km~s$^{-1}$~d$^{-1}$ and $-77\pm16$~km~s$^{-1}$~d$^{-1}$ for the phase range of +7 to +25 days from peak $B$-band light.

For SNe Ia spectra near peak light, both \textit{p}W regions are visible, however, a couple weeks later Si \textsc{ii} $\lambda$5972 (\textit{p}W6) is no longer detectable, but Si \textsc{ii} $\lambda$6355 is, as in the case of SN~2016geu. The classifications above are based on the strength of the \textit{p}W features at peak light, whereas the earliest epoch investigated here corresponds to $t_{B}^{\rm max}$=+7.9~days (from \citealt{Goobar17}).  Inspection of fig. 11 from \citet{Folatelli13} shows that, for a given SN Ia, the value of \textit{p}W7 is approximately the same for the 5$-$10 days either side of $t_{B}^{\rm max}$, albeit with a lot of scatter.  If the \textit{p}W7 values at this first epoch from the lens-subtracted/unsubtracted spectra ($65\sim70$~\AA) are a suitable proxy for its value at peak light, then they imply SN~2016geu can be classified as either SS or CN.  It is clearly not a BL SN Ia, and as the CL denomination is based purely on \textit{p}W6 only, we cannot comment on its CL nature.  In terms of the velocity gradient, SN~2016geu is most consistent with the CN sub-class (roughly 2$\sigma$ larger), and is clearly inconsistent with the BL, CL and SS sub-classes. In general, SN~2016geu is consistent with the CSP combined sample within 1$\sigma$, and in terms of both observables, SN~2016geu can be considered as a normal SN Ia.

\subsection{Wang et al. (2009)}

\citet{Wang09} investigated a sample of $N=158$ SNe Ia near peak $B$-band light, and classified their SNe into two sub-classes: normal and high-velocity (HV).  These two groups were based on the Si~\textsc{ii}~$\lambda$6355 velocity at peak $B$-band light, where the samples had averages of $\left < v \right > = 10600\pm400$~km~s$^{-1}$ and $\left < v \right > \ge 11800$ ~km~s$^{-1}$, respectively.  \citet{Wang09} find that the HV group has narrower $M_B$ and $\Delta m_{15,B}$ distributions than the normal group, and there is overlap in phenomenological properties of the HV group with the HVG group of \citet{Benetti05}.  These authors find that the normal group has a velocity evolution after peak $B$-band light of $\left < \dot{v} \right > =\approx -40$~km~s$^{-1}$~d$^{-1}$, which is similar to that measured for the LVG of \citet{Benetti05}.

The value of $v_{B}^{\rm max} = 11,950 \pm 140$~km~s$^{-1}$ measured for SN~2016geu places it in the HV group, and is more than 3$\sigma$ more rapid than the normal group average.

\begin{figure*}
 \centering
 \includegraphics[bb=0 0 1440 720,scale=0.35,keepaspectratio=true]{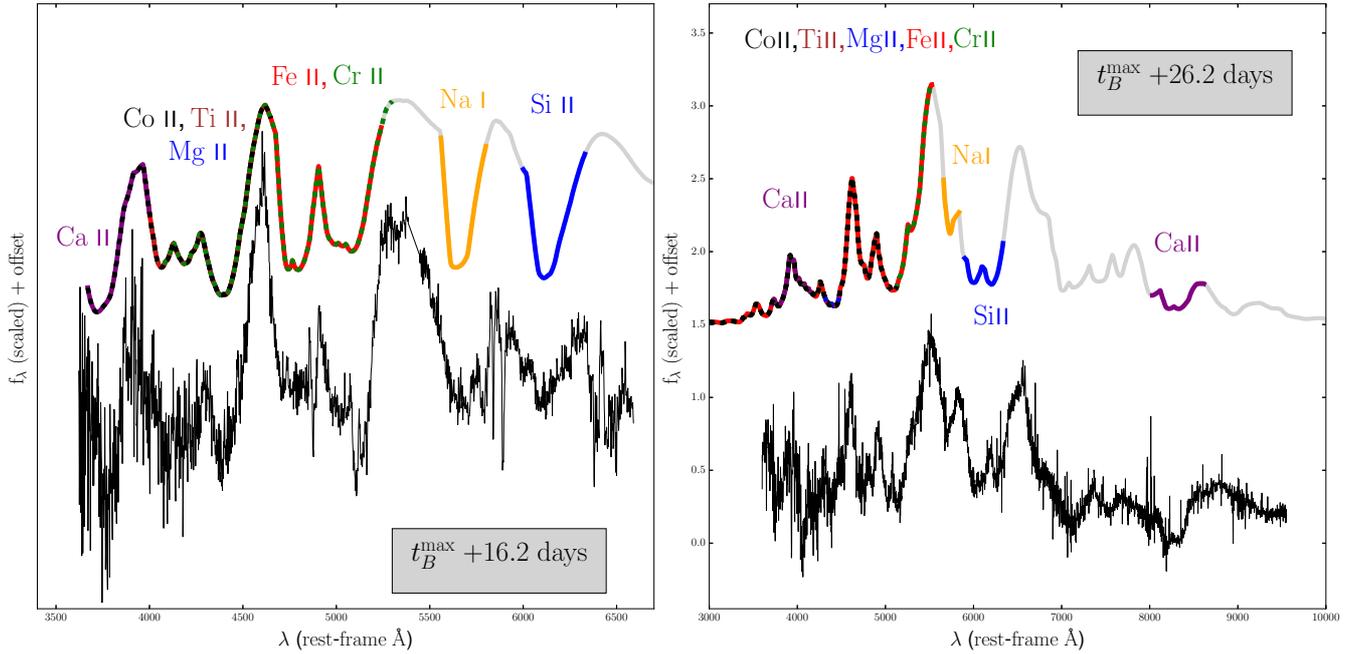}
 \caption{Comparison of the rest-frame spectra with synthetic SYN++ spectra.  \textit{Left}: GTC spectrum (in black) from $t_{B}^{\rm max} +16.2$~days.  The various transitions are colour-coded to show the wavelength intervals where each element/ion can be attributed to specific absorption features. \textit{Right}: X-shooter spectrum (black) from $t_{B}^{\rm max} +26.2$~days, which has been mildly smoothed to aid clarity.  As before, telluric regions have been removed.  The parameters of the synthetic spectra are given in Table \ref{table:SYN}.}
 \label{fig:syn}
\end{figure*}

\begin{table*}
\centering
\setlength{\tabcolsep}{4.0pt}
\setlength{\extrarowheight}{3pt}
\caption{SN 2016geu: SYN++ model parameters}
\label{table:SYN}
\begin{tabular}{cccccccc}
\hline	
$\Delta t_{B}^{\rm max}$ (d)$^\dagger$	&	Element/Ion	&	$T_{\rm exc}$ (K)	&	$T_{\rm BB}$ (K)	&	$v_{\rm ph}$ (km s$^{-1}$)	&	log $\tau$	&	$v_{\rm max}$ (km s$^{-1}$)	&	wavelength range(s) (\AA)	\\
\hline
+16.2	&	Na \textsc{i}	&	7000	&	7000	&	10,000	&	0.3	&	18.0	&	5560:5800	\\
+16.2	&	Mg \textsc{ii}	&	7000	&	7000	&	10,000	&	0.4	&	14.0	&	4290:4450	\\
+16.2	&	Si \textsc{ii}	&	7000	&	7000	&	10,000	&	0.5	&	16.5	&	3660:3900; 6000:6330	\\
+16.2	&	Ca \textsc{ii}	&	7000	&	7000	&	10,000	&	0.5	&	22.0	&	3670:4000	\\
+16.2	&	Ti \textsc{ii}	&	7000	&	7000	&	10,000	&	0.3	&	12.0	&	3500:4500	\\
+16.2	&	Cr \textsc{ii}	&	7000	&	7000	&	10,000	&	2.0	&	13.0	&	4080:5300	\\
+16.2	&	Fe \textsc{ii}	&	7000	&	7000	&	10,000	&	1.0	&	16.0	&	3975:5240	\\
+16.2	&	Co \textsc{ii}	&	7000	&	7000	&	10,000	&	0.3	&	12.0	&	3480:4650	\\
\hline															
+26.2	&	Na \textsc{i}	&	7000	&	6200	&	8000	&	-0.2	&	14.0	&	5660:5875	\\
+26.2	&	Mg \textsc{ii}	&	7000	&	6200	&	8000	&	1.0	&	12.0	&	4310:4485	\\
+26.2	&	Si \textsc{ii}	&	7000	&	6200	&	8000	&	-0.5	&	20.0	&	3640:3860; 5990:6335	\\
+26.2	&	Ca \textsc{ii}	&	7000	&	6200	&	8000	&	3.0	&	14.0	&	3710:4150; 8030:8615	\\
+26.2	&	Ti \textsc{ii}	&	7000	&	6200	&	8000	&	0.5	&	13.0	&	3000:5160	\\
+26.2	&	Cr \textsc{ii}	&	7000	&	6200	&	8000	&	3.0	&	13.0	&	3000:5525	\\
+26.2	&	Fe \textsc{ii}	&	7000	&	6200	&	8000	&	2.0	&	16.0	&	3000:5525	\\
+26.2	&	Co \textsc{ii}	&	7000	&	6200	&	8000	&	0.3	&	12.0	&	3000:5130	\\
\hline					
\end{tabular}

\end{table*}

\section{Spectral modelling with SYN++}
\label{sec:synthetic}

In Section \ref{sec:SNIa_compare}, visual comparison of the GTC and X-shooter spectra with those of well-observed, nearby SNe Ia 1994D and 2011fe revealed the presence of iron-peak elements and intermediate-mass elements.  To gain a deeper understanding of which elements in the SN ejecta may be responsible for the various absorption features in the spectra of SN~2016geu, we created synthetic SN spectra using SYN++ \citep{Thomas11} for two of the spectra presented here: $t_{B}^{\rm max} +16.2$ and $+26.2$~days.  The original and synthetic spectra are presented in Fig. \ref{fig:syn}, while the best-fitting parameters are given in Table \ref{table:SYN}.

In the left-panel of Fig. \ref{fig:syn} are the GTC spectra from $t_{B}^{\rm max} +16.2$.  When creating the synthetic SYN++ spectra, we assumed an excitation temperature of $T_{\rm exc}=7000$~K following the SYNOW modelling performed by \citet{Branch08} for SNe Ia a similar post-maximum epoch.  The photospheric velocity ($v_{\rm ph}$), the blackbody temperature ($T_{\rm BB}$), the density of each element/ion (log $\tau$) and its maximum velocity ($v_{\rm max}$) were manually changed to produce a good visual fit.  Following \citet{Parrent12}, we used a constant term of 1.40 and a quadratic warp term of -1.9, to get the synthetic spectrum to match the luminosity of the GTC spectrum, and to obtain good agreement between the synthetic spectrum and the observations at the blue end. For the X-shooter spectrum, we used a constant term of 0.9, and a linear term of -1.2.

As seen visually in Fig. \ref{fig:2_SN_compare}, the elements/ions responsible for the main absorption features are (in increasing atomic number): Na~\textsc{i}, Mg~\textsc{ii}, Si~\textsc{ii}, Ca~\textsc{ii}, Ti~\textsc{ii}, Cr~\textsc{ii}, Fe~\textsc{ii}, and Co~\textsc{ii}.  Si~\textsc{ii} is, as expected, most prominent in the wavelength range $6000 \lambda \le 6330$~\AA, while also contributing some of the absorption seen between $3600-3900$~\AA.  Much of the absorption seen between $4000-5300$~\AA~can be attributed to Fe~\textsc{ii}, with minor contributions from Cr~\textsc{ii} over the same wavelength interval.  Ti ~\textsc{ii} also contributes to absorption between $3500-4650$~\AA, in particular the absorption line near 4190~\AA.  Na~\textsc{i} can be attributed to the feature between $5560-5800$~\AA, while the absorption bluewards of 4000~\AA~is due to Ca~\textsc{ii} (H\&K).  Modest absorption due to  Mg~\textsc{ii} is seen between $4290-4450$~\AA, and Co~\textsc{ii} between $3500-4650$~\AA, which the most prominent absorption near $\approx 4000$~\AA.  The densities of each element/ion are similar (log$\tau$=-0.1 to -0.5), however that of Cr~\textsc{ii} is significantly larger (log$\tau=2.0$).  The maximum velocity of each element/ion ranges from 12,000~km~s$^{-1}$ for Co~\textsc{ii} to 22,000~km~s$^{-1}$ for Ca~\textsc{ii}.  The maximum velocity of Fe~\textsc{ii} was found to be 16,000 km~s$^{-1}$; larger velocities resulted in too much line blending in the synthetic spectra.

The X-shooter spectrum from $t_{B}^{\rm max} +26.2$ is shown in the right-panel of Fig. \ref{fig:syn}, which spans a wider range in wavelength than the earlier GTC spectrum.  This time we are able to investigate both the blue \textit{and} red Ca~\textsc{ii} absorption features.  As before, we kept the excitation temperature at 7000~K.  The BB temperature was 6200~K, and the photospheric velocity was found to be 8000~km~s$^{-1}$.  Given that only 10 days had passed since the GTC epoch, the elements and ions likely responsible for the various absorption features are similar to the previous epoch.   Indeed the relative densities and maximum velocities are comparable between the two epochs, though the log($\tau$) values are generally smaller in this epoch. The strongest transitions here are Ca~\textsc{ii}, Cr~\textsc{ii} and Fe~\textsc{ii}.

\section{Discussion \& Conclusions}
\label{sec:conclusion}


In this work we investigated the spectroscopic properties of the gravitationally lensed SN Ia 2016geu.  A total of six epochs of long-slit spectra were obtained with GTC-OSIRIS and VLT-X-shooter, at phases ranging from +16.2 to +59.4 days after peak $B$-band light.  Using the light curves obtained by \citet{Goobar17} and modelled with SALT2, we estimated and removed the contribution of the lens galaxy from our spectra.  The resultant spectra were then dominated by the SN itself, though a small, but unknown, portion of the light was contaminated by its host galaxy.  However, that the modelled \textit{p}W values obtained from the lens-subtracted spectra matched well with those from the CSP sample \citep{Folatelli13}, see Fig.~\ref{fig:4_pWs}, leads us to suggest that the host contamination is only minimal; the removal of an additional spectrum will lower the overall continuum of the subtracted spectra, resulting in larger \textit{p}W values.  Indeed, if the \textit{p}W3 and \textit{p}W4 values increased any further, they would be in excess of the largest values measured in the CSP sample.  

We compared two of these epochs (+16.2 and +26.2~days) with nearby SNe Ia 1994D and 2011fe.  SN~1994D has been classified as a normal, CN and LVG SN Ia, while SN~2011fe as a normal and CN, and on the cusp between the LVG and HVG group classifications.  Visual comparison with the nearby SNe Ia revealed line-transitions corresponding to several iron-peak and intermediate-mass elements, which are typical of post-maximum SNe Ia at +2 to +4 weeks.

Next, we classified SN~2016geu using the schema of \citet{Benetti05}, \citet{Branch06} and \citet{Wang09}:

\begin{itemize}
 \item \citet{Benetti05}: SN~2016geu is most consistent with the HVG group, and inconsistent with the FAINT group.
 \item \citet{Branch06}: SN~2016geu is most consistent with the CN group, and inconsistent with the BL, SS and CL classes.
 \item \citet{Wang09}: SN~2016geu is most consistent with the HV group, and more than $3\sigma$ more rapid than the normal sub-class.
\end{itemize}

It is interesting to note here that the CN SNe Ia in the CSP sample (\citealt{Folatelli13}) are not HV objects -- indeed there are very few cases where a SN Ia has been classified as both a CN and HV (e.g., SNe 2006is and 2005ku, both in the CSP sample). Instead, most of the HV objects in their sample belong to the BL class.  

When fitting the \textit{p}Ws, we found that both the Fe~\textsc{ii} and Mg~\textsc{ii} \textit{p}Ws increased sharply around $t_{\rm B}^{\rm max} + 15$~days.  This behaviour was also observed in the sample investigated by \citet{Folatelli13}.  The Si~\textsc{ii}~$\lambda6355$ \textit{p}W increased steadily over the investigated time-period, while the Ca~\textsc{ii} H\&K \textit{p}W remained constant.  The strength of a given \textit{p}W can be used as a proxy for how much of a given element is present at or above the photosphere at the observation time.  Over time, the SN ejecta expands and cools, and the observer is offered a deeper and deeper view of the SN ejecta.  Thus, if the value of a given \textit{p}W increases as the velocity decreases (and hence time increases), this can imply a more central location of the element/ion. Thus, the measured increase in the \textit{p}Ws associated with Fe~\textsc{ii} and Mg~\textsc{ii} may imply a more central location for a sizable amount of these species. In contrast, the smoother \textit{p}W evolution of  Si~\textsc{ii}~$\lambda6355$ implies a more mixed and homogeneous distribution.  We note however that this conclusion is rather simplified, and it does not consider the effects arising from temperature and the recombination rates of various elements in the ejecta.  As such, while an increase of a given \textit{p}W measurement of a given element/ion can imply its increased abundance, temperature can affect the recombination rate which in turn also affects the strength of the measured \textit{p}Ws  Moreover, as the Fe and Mg \textit{p}Ws are affected by blending, their measured changes can arise from changes in the blending of the various transitions giving rise to the specific \textit{p}W, rather than an increase in its relative abundance.  Thus, the inferred location of the Fe, Mg and Si in the SN ejecta needs to be interpreted with these caveats in mind.


Next, we created synthetic spectra using the 1D SYN++ code \citep{Thomas11} for the two aforementioned GTC and X-shooter spectra.  As from the visual comparison with the nearby SNe Ia, the main elements and ions needed to reproduce the observations were iron-peak elements (Ti ~\textsc{ii}, Cr~\textsc{ii}, Fe~\textsc{ii}, and Co~\textsc{ii}) and intermediate-mass elements (Na~\textsc{i}, Mg~\textsc{ii}, Si~\textsc{ii}, and Ca~\textsc{ii}).  The two investigated epochs were 10 rest-frame days apart, and we found that the main elements, and their relative strengths (densities) were comparable for the two epochs.  Notably, the log($\tau$) values of Ca~\textsc{ii}, Cr~\textsc{ii} and Fe~\textsc{ii} increased in the second epoch, implying their increased abundance.  This increase also appears to correspond with the increase in their measured \textit{p}W values. 

Based on the above classifications and modelling, we can conclude with some confidence that SN~2016geu is a normal SN Ia, and it possesses properties similar to SNe Ia in the local universe.  This result supports the conclusions of \citet{Hook05} and \citet{Petrushevska17}, which suggest there is very little, if any, difference in the observational properties of SNe Ia up to $z\approx0.4$, as shown here, and perhaps even up to $z\approx1.4$, as shown by the latter authors.  We note that the redshift of SN~2016geu is at the lower-end of the redshift range of SNe Ia investigated by \citet{Maguire12}, and hence we cannot comment on the slight redshift-variance observed in their sample.  Inevitably more moderate to high-$z$ SNe Ia are required to further test the assumption of the invariance of their observational properties, the conclusions of which will have severe implications on their continued use in SN-cosmology.

\section*{Acknowledgments}

ZC acknowledges support from the Juan de la Cierva Incorporaci\'on fellowship IJCI-2014-21669.  AdUP acknowledges support from funding associated to a Ram\'on y Cajal fellowship RyC-2012-09975.  DAK acknowledges support from the Juan de la Cierva Incorporaci\'on fellowship IJCI-2015-26153.  ZC, DAK \& AdUP acknowledge support from the Spanish research project AYA 2014-58381-P.  CG was funded by the Carlsberg Foundation. 

This work is based on observations made with the GTC telescope, in the Spanish Observatorio del Roque de los Muchachos of the Instituto de Astrof\'isica de Canarias, under the Director's Discretionary Time (proposal ID: GTC2016-058).

\bsp

\label{lastpage}

\end{document}